\documentclass[12pt]{article} 
\usepackage{graphicx,amsmath,amsthm,amssymb}
\begin{document}
\title{Linear noise approximation of noise-induced oscillation in NF-$\kappa$B signaling network}
\author{Jaewook Joo\\
Department of Physics and Astronomy, University of Tennessee, Knoxville, \\
TN, 37996, USA}
\maketitle

\begin{abstract}

NF-$\kappa$B, one of key regulators of inflammation, apoptosis, and differentiation, was found to have noisy oscillatory shuttling between the nucleus and the cytoplasm in single cells when cells are stimulated by cytokine TNF$\alpha$. We present the analytical analysis which uncovers the underlying physical mechanisms of this spectacular noise-induced transition in biological networks. Starting with the master equation describing both signaling and transcription events in NF-$\kappa$B signaling network, we derived the macroscopic and the Fokker-Planck equations by using van Kampen's sysem size expansion. Using the noise-induced oscillatory signatures present in the power spectrum, we constructed the two-dimensional phase diagram where the noise-iduced oscillation emerges in the dynamically stable parameter space. 

\end{abstract}

\section{Introduction}

Oscillations are prevalent in biology. Neurons periodically fire pulses, 
cells undergo cell-cycle, and living organisms maintain their internal biological 
clock of about 24 hours (Murray 1980). Strikingly, live cell imaging techniques revealed that 
biochemical reactions are likely to exhibit oscillatory behavior in single cells. For example, 
perpetual oscillations of a tumor suppressor protein p53 and its suppressor mdm2 have been observed in single cells when cells were exposed to ionizing UV irradiation (Lahav et al 2004). In addition, NF-$\kappa$B, one of key regulators of inflammation, apoptosis, and differentiation, was found to shuttle between the nucleus and the cytoplasm of single cells when cells are stimulated by cytokine TNF$\alpha$ (Nelson et al 2004). Despite its universal appearance and its potential biological functions, the underlying physical mechanisms of such biological oscillations in noisy cellular environments have been neither discussed nor investigated to a mathematically satisfactory degree. Sadly, most of publicly available investigations on the subject are based on numerical simulations and are quite limited in providing physical and mathematical insights behind these fascinating natural phenomena. In this manuscript, we'd like to present the analytical analysis which uncovers the underlying mechanisms of the spectacular noise-induced  transition in one of biochemical oscillations.

In biological, physical,and chemical systems, stochastic fluctuations (white noise) can induce
oscillations in a dynamical system with an external
periodic forcing (Steuer 2004; Pikovsky et al 1997; Neiman et al 1997; Gammaitoni et al 1998; Benzi et al 1982). This phenomena
are called stochastic resonance where the background white noise
amplifies the external periodic input signal and such an
amplification is optimized at an intermediate level of noise. On
the other hand, noise can also coherently induce oscillations
without external periodic modulation. In this case the deterministic
dynamical system has a stable fixed point, but has a long excursion
in phase trajectory because of nearby dynamical instability; the
system is constantly kicked off a stable fixed point by stochastic
fluctuations and follows a deterministic phase trajectory at the
natural frequency at which the system reaches its stable fixed
point. This coherent noise-induced oscillation (NIO) was investigated in several model systems
(Wiesenfeld 1985; Gang et al 1993; Mckane \& Newman 2007). 
In addition, researchers revisited the previous studies of biological
oscillations and investigated the effects of noise on the
existence \& emergence of limit cycles. The deterministic limit
cycle emerges at some confined parameter space whereas stochastic
fluctuations increase such a domain where noise not only stabilizes
the limit cycle, but also induces oscillations at a parameter space
where the deterministic dynamics produces a stable dynamics.

We are interested in the effects of noise on the
stable dynamics generated by negative feedback loops that are most
commonly found in the biological networks. In the recent paper
(Joo et al 2010), the authors presented NIO dynamic phenomena on a
large-scale signal transduction network of NF-$\kappa$B. A full stochastic
two-compartmental model of the NF-$\kappa$B signal transduction network
consisting of 28 biochemical species and 71 kinetic parameters was
investigated with the help of Gillespie stochastic simulations
and, depending on the choice of the kinetic parameter values,
yielded a phase diagram consisting of oscillations
and damped-oscillations of NF-$\kappa$B shuttling in carefully chosen
two-dimensional parameter space. The authors successfully demonstrated the
noise-induced oscillations of the NF-$\kappa$B shuttling and identified
the inherent stochastic fluctuations as the potential source of those
fantastic stochastic phenomena. However, the full scale stochastic simulations
with the large-scale size network and its nonlinear complexity,
put a hard limit on our capability in investigating the
characteristics of the bifurcation, the dependence of the NIO on
the remaining parameter space. What is worse, the numerical
analysis suffers from the lack of kinetic details, e.g.,
functional forms of reactions and associated kinetic parameter
values. Thus, to overcome the difficulties described above, we
like to perform stochastic analysis of a reduced network of NF-$\kappa$B,
assuming a linear response of the system to small perturbation

In this paper, using the stochastic reduced order
model of NF-$\kappa$B signaling network and linear noise approximation, we present the underlying physical mechanisms of the NIO and show analytically the emergence of the NIO. 
We took into account the full stochastic nature of transcriptional processes in the model. 
We started with the master equation describing both signaling and transcription events, and then derived the macroscopic and Fokker-Planck equations by using van Kampen's sysem size expansion. First, we performed linear stability analysis of macroscopic equations. Second, using the NIO signatures present in the power spectrum derived from Langevin equations, we constructed the two-dimensional phase diagram where the NIO emerges in the dynamically stable parameter space. 

\section{Model description}

\begin{figure}[htp]
\centering
\includegraphics[height=10cm,angle=0]{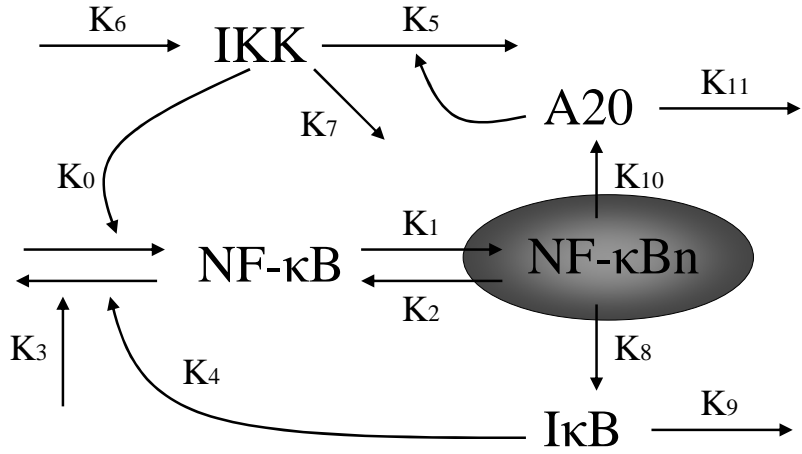}
\caption{Minimal model consisting of five proteins, IKK, cytoplasmic
NF-$\kappa$B (NF-$\kappa$B), nuclear NF-$\kappa$B (NF-$\kappa$Bn), I$\kappa$B, and A20. The synthesis of both A20 and I$\kappa$B depends on the NF-$\kappa$Bn. $K$'s denote
the reaction rates of the respective reactions. The external signal
is received from the $K_6$.} \label{fig1}
\end{figure}

The full network consisting of 28 species and 71 kinetic parameters 
can be topologically reduced to an
intermediately reduced network consisting of seven components, by
using sensitivity analysis that calculates the correlation coefficients 
between kinetic parameters and the NF-$\kappa$B
dynamic response and eliminates insignificant kinetic
parameters whose change doesn't significantly affect the NF-$\kappa$B response
as presented in Ref. (Joo et al 2007). We further
eliminate two fast variables, mRNA I$\kappa$B and mRNA A20, whose half
life are less than a half hour, obtaining a reduced network as shown in Fig.~\ref{fig1}.

In more rigorous study of the stochastic dynamical system, one should 
start with a stochastic model of full NF-$\kappa$B signaling network and 
adiabatically eliminate the "fast varying variables".  However, the adiabatic elimination of the fast varying variables from the Makovian stochastic system leads to non-Markovian system, 
making analysis almost impossible and thus adiabatic 
elimination techniques  are limited only to a small stochastic system. 
Thus, we take an non-rigorous but practical way of model reduction; 
we reduce the complexity or the dimension of the NF-$\kappa$B signaling 
network by using various heuristic techniques and then the reduced 
network is converted to a stochastic minimal model.

The current stochastic model neglects an intermediate step of protein synthesis, 
namely mRNA, that a typical stochastic model with modest level of 
complexity always contains. But, the mRNA and protein synthesis 
processes are too complicated to model them with a finite number of steps. Thus, we simply 
represent them with a single step (a single variable). Lastly, we include the stochastic fluctuations in 
DNA and protein NF-$\kappa$ interaction.

The dynamic system under our consideration regulates the shuttling dynamics of NF-$\kappa$B protein
between cytoplasm and nucleus. In the absence of stimulus, the
activated IKK is absent and NF-$\kappa$B can exist either in a free form
(NF-$\kappa$B) or can be bound with inhibitor $\kappa$B, forming a complex
(I$\kappa$B:NF-$\kappa$B). The total amount of NF-$\kappa$B, $X_o$, is conserved,
enabling us to omit the expression of a protein complex
I$\kappa$B:NF-$\kappa$B from our model because the amount of protein complex
I$\kappa$B:NF-$\kappa$B is simply $X_o$-NF-$\kappa$ B. The free NF-$\kappa$B can enter into
nucleus with a rate $k_1$ or moves out of nucleus with a rate
$k_2$ by nuclear transport proteins. The liberated \& translocated
NF-$\kappa$B into nucleus initiates mRNA \& protein syntheses of I$\kappa$B and
A20 with rates $k_8$ and $k_{10}$ respectively, both of which forms
negative feedback loops, negatively regulating the translocation
of NF-$\kappa$B into nucleus. The newly synthesized I$\kappa$B bind to NF-$\kappa$B
with a rate $k_4$, forming a protein complex I$\kappa$B-NF-$\kappa$B. Upon
stimulus, I$\kappa$B kinase (IKK) is activated with a constant rate, $S
\cdot k_6$ where $S$ is stimulus strength and also is deactivated
with a rate $k_7$. This IKK induces I$\kappa$B decay from I$\kappa$B-NF-$\kappa$B and
liberates NF-$\kappa$B with a rate $k_o$, which translocates NF-$\kappa$B into
nucleus, increasing the synthesis of negative regulators I$\kappa$B and
A20. In turn, A20 inactivates the activated IKK with a rate $k_5$.
A20 and I$\kappa$B also decay with rates $k_9$ and $k_{11}$ respectively.

\section{Derivation of chemical master equation}
We intend to use a probabilistic model of the NF-$\kappa$B system to be able
to capture unusual stochastic phenomena such as noise-amplified
oscillation. Let us define the joint probability density
$P_t(\underline{\eta})$denoting the probability of the system to be
in a state $\underline{\eta}(t)= (C_t, D_t, X_t, Y_t, Z_t)$ at time
$t$ where $C_t$, $D_t$, $X_t$, $Y_t$, and $Z_t$ denote the molecule
number of $I\kappa B$, A20, NF-$\kappa$B, $NF\kappa Bn$, and IKK proteins at
time $t$, respectively. The time evolution of the joint probability
is determined by the transition probability per unit time
$T(\underline{\eta} | \underline{\eta'};t)$ of going from a state
$\underline{\eta}$ to a state $\underline{\eta'}$. We assume that
the transition probabilites do not depend on the history of the
previous states of the system but only on the immediately past
state. There are only a few transitions that are allowed to take
place and their transition rates are
\begin{eqnarray}
\label{transition} T_0(..., X_t,... |..., X_t+1,...;t) &=& k_o
\Omega_C (\frac{X_o-X_t-Y_t}{\Omega_C})(\frac{Z_t}{\Omega_C}),
\\ \nonumber
T_1(..., X_t, Y_t,... |..., X_t-1, Y_t+1, ...;t) &=& k_1 \Omega_C
(\frac{X_t}{\Omega_C}),
\\ \nonumber
T_2(..., X_t, Y_t,... |..., X_t+1, Y_t-1, ...;t) &=& k_2 \Omega_N
(\frac{Y_t}{\Omega_N}),
\\ \nonumber
T_3(..., X_t,... |..., X_t-1,...;t) &=& k_3 \Omega_C
(\frac{X_t}{\Omega_C}),
\\ \nonumber
T_4(C_t,..., X_t,... | C_t,..., X_t-1,...;t) &=& k_4 \Omega_C
(\frac{X_t}{\Omega_C})(\frac{C_t}{\Omega_C}),
\\ \nonumber
T_5(..., D_t, ..., Z_t |..., D_t,..., Z_t-1;t) &=& k_5 \Omega_C
(\frac{D_t}{\Omega_C}(\frac{Z_t}{\Omega_C}),
\\ \nonumber
T_6(..., Z_t |..., Z_t+1;t) &=& k_6 S \Omega_C,
\\ \nonumber
T_7(..., Z_t |..., Z_t-1;t) &=& k_7 \Omega_C
(\frac{Z_t}{\Omega_C}),
\\ \nonumber
T_8(C_t,..., Y_t,... | C_t+1,..., Y_t,...;t) &=& k_8 \Omega_C
(\frac{Y_t}{\Omega_N}),
\\ \nonumber
T_9(C_t,... | C_t-1,...) &=& k_9 \Omega_C (\frac{C_t}{\Omega_C}),
\\ \nonumber
T_{10}(..., D_t,..., Y_t,... |..., D_t+1,..., Y_t,...;t) &=&
k_{10} \Omega_C (\frac{Y_t}{\Omega_N}),
\\ \nonumber
T_{11}(..., D_t,... | ...,D_t-1,...;t) &=& k_{11} \Omega_C
(\frac{D_t}{\Omega_C})
\\ \nonumber
\end{eqnarray}
where $\Omega_C$ and $\Omega_N$ are the cytoplasmic and nuclear
volumes of a typical cell, respectively. $X_o$ is an initial and
conserved molecule number of NF-$\kappa B$ protein. The above
transition rates are provided as defined in Ref. (van Kampen 2001) and
given in three terms in the following order: a reaction rate, a
reactive volume where the reaction takes place, and the
concentration(s) of all involved components, e.g., $k_1 \cdot
\Omega_C \cdot \frac{X_t}{\Omega_C}$. All the above transitions
occur at the cytoplasm except the transition $T_2$. The transition
$T_2$ (or $T_1$) describes a binary (associative) reaction between
a NF-$\kappa$B protein in nucleus (or in cytoplasm) and a transport protein
called ``exportin'' present in nucleus (or ``importin'' present in
cytoplasm), leading to the transportation of NF-$\kappa$B protein into cytoplasm
(or into nucleus). The reactive volume is thus nuclear volume $\Omega_N$
(or cytoplasmic volume $\Omega_C$) and assuming a constant concentration
of importin and exportin being independent of the system's variables,
their concentrations are absorbed into reaction rates. Two transitions,
$T_8$ and $T_{10}$, describe a complex mRNA and protein synthesis in
a single term and those syntheses begin from the reaction between
nuclear NF-$\kappa$B and DNA in nucleus and ends at the production of a
protein out of Ribosome in cytoplasm. Thus, the reactive volume of
these two transitions is the cytoplasmic volume. In fact, these two transitions 
oversimplify very stochastic nature of transcription and translation. A more realistic model 
is provided and analyzed in appendix A.

The stochastic process specified by the transition rates in 
Eq.~(\ref{transition}) is Markovian, thus we can immediately write down 
a master equation governing the time evolution of the joint probability 
$P_t(\underline{\eta})$. The rate of change of the joint probability 
$P_t(\underline{\eta})$ is the sum of transition rates from all 
other states $\underline{\eta'}$ to the state $\underline{\eta}$, 
minus the sum of transition rates from the state $\underline{\eta}$ 
to all other states $\underline{\eta'}$:
\begin{eqnarray}
\label{master_equation}
\frac{d P_t (\underline{\eta})}{dt} &=&
(E^{-1}_X-1)T_0(..., X_t,... |..., X_t+1,...;t)P_t (\underline{\eta})
\\ \nonumber
&+& (E_X E^{-1}_Y-1) T_1(..., X_t, Y_t,... |..., X_t-1, Y_t+1, ...;t)P_t (\underline{\eta})
\\ \nonumber
&+& (E^{-1}_X E_Y -1) T_2(..., X_t, Y_t,... |..., X_t+1, Y_t-1, ...;t)P_t (\underline{\eta})
\\ \nonumber
&+& (E_X-1) T_3(..., X_t,... |..., X_t-1,...;t)P_t (\underline{\eta})
\\ \nonumber
&+& (E_X E_C-1) T_4(C_t,..., X_t,... | C_t-1,..., X_t-1,...;t)P_t
(\underline{\eta})
\\ \nonumber
&+& (E_Z-1) T_5(..., D_t, ..., Z_t |..., D_t,..., Z_t-1;t)P_t (\underline{\eta})
\\ \nonumber
&+& (E^{-1}_Z-1) T_6(..., Z_t |..., Z_t+1;t)P_t (\underline{\eta})
\\ \nonumber
&+& (E_Z-1) T_7(..., Z_t |..., Z_t-1;t)P_t (\underline{\eta})
\\ \nonumber
&+& (E^{-1}_C-1) T_8(C_t,..., Y_t,... | C_t+1,..., Y_t,...;t)P_t (\underline{\eta})
\\ \nonumber
&+& (E_C-1) T_9(C_t,... | C_t-1,...)P_t (\underline{\eta})
\\ \nonumber
&+& (E^{-1}_D-1) T_{10}(..., D_t,..., Y_t,... |..., D_t+1,..., Y_t,...;t)
P_t (\underline{\eta})
\\ \nonumber
&+& (E_D-1) T_{11}(..., D_t,... | ...,D_t-1,...;t)P_t (\underline{\eta})
\end{eqnarray}
where $E^{\pm}_{\alpha}$ is a step operator which acts on any function 
of $\alpha$ according to $E^{\pm}_{\alpha}f(\alpha\pm1,...)=f(\alpha\pm1,...)$.

\section{System size expansion of the chemical master equation}
In this section we will apply van Kampen's elegant ``system size
$\Omega$-expansion method'' to the chemical master equation in
Eq.~(\ref{master_equation}). This van kampen's $\Omega$-expansion
method not only allows us to obtain a deterministic version of the
stochastic model in Eq.~(\ref{master_equation}) but also enables us
to find Gaussian corrections to the deterministic result. We choose
two system sizes, cytoplasmic volume $\Omega_C$ and nuclear volume
$\Omega_N$, and expand the master equation in order of
$\Omega^{1/2}_C$ and $\Omega^{1/2}_N$. A required condition for
valid use of $\Omega$-expansion is the stability of fixed points and restricts us to perform this analysis 
only inside of the dynamically stable region.

For large and finite cytoplasmic and nuclear volume, $\Omega_C$ and
$\Omega_N$, we would expect $P (\underline{\eta}) $ to have a finite
width of order $\Omega^{1/2}$. The variables (X, Y, C, D, Z) are
stochastic and we introduce new stochastic variables ($\xi_X$,
$\xi_Y$, $\xi_C$, $\xi_D$, $\xi_Z$): $X=x \Omega_{C} +
\sqrt{\Omega_C} \xi_{X}$, $Y=y\Omega_{N}+\sqrt{\Omega_N} \xi_{Y}$,
$C=c \Omega_{C} + \sqrt{\Omega_C} \xi_{C}$, $D=d \Omega_{C} +
\sqrt{\Omega_C} \xi_{D}$, and $Z=z \Omega_{C} + \sqrt{\Omega_C}
\xi_{Z}$. These new stochastic variables represent inherent noise
contribute to deviation of the system from the macroscopic dynamical
behavior.

The new joint probability density function $\Pi_t$ is defined by
$\Pi_t (\underline{\xi}) \equiv P_t (\underline{\eta})$. Let us
define the step operators $E^{\pm}_{\alpha}$ which change the state
$\alpha$ to $\alpha+1$ and therefore $\xi_{\alpha}$ into
$\xi_{\alpha}+\Omega^{1/2}$, so that the continuous approximation of
the step operator is
\begin{equation}
E^{\pm}_{\alpha}=1 \pm \Omega^{-\frac{1}{2}}
\frac{\partial}{\partial \xi_{\alpha}} +\frac{\Omega^{-1}}{2}
\frac{\partial^{2}}{\partial \xi^{2}_{\alpha}} \pm ...
\end{equation}
where $\Omega$ becomes $\Omega_C$ for cytoplasmic component $\alpha$
and $\Omega_N$ for nucleoplasmic $\alpha$. The time derivative of
the the joint probability $P_t(\underline{\eta})$ in
Eq.~(\ref{master_equation}) is taken at a fixed state
$\underline{\eta}$, which implies that the time-derivative on both
sides of $\alpha=\Omega \alpha + \Omega^{1/2} \xi_{\alpha}$ should
lead to $d \xi_{\alpha}/dt=-\Omega^{1/2}$  $d \alpha /dt$.
Therefore,
\begin{equation}
\label{omega_expansion} \frac{d
P_t(\underline{\eta})}{dt}=\frac{\partial
\Pi_t(\underline{\xi})}{\partial t} -\sum_{\alpha=X,C,D,Z} \{
\Omega^{1/2}_C \frac{d \alpha}{dt} \frac{\partial \Pi_t
(\underline{\xi})}{\partial \xi_{\alpha}} \} -\Omega^{1/2}_N \frac{d
y}{dt} \frac{\partial \Pi_t (\underline{\xi})}{\partial \xi_{Y}}
\end{equation}
The initial condition for the above parabolic PDE is
$P_{t=0}(\underline{\eta})=\delta_{\underline{\eta},
\underline{\eta_0}}$. When the Eq.~(\ref{master_equation}) is
expanded in order of $\Omega^{1/2}_C$ and $\Omega^{1/2}_N$, we can
collect several powers of $\Omega^{1/2}_C$ and $\Omega^{1/2}_N$. The
macroscopic equations emerge from the terms of order
$\Omega^{1/2}_C$ and $\Omega^{1/2}_N$ and the linear Fokker-Planck
equation is obtained from the terms of order $\Omega^0_C$ and
$\Omega^0_N$.

\section{Emergence of the macroscopic equations}
By matching the terms in order of $\Omega^{1/2}$ in
Eqs.~(\ref{master_equation}) and ~(\ref{omega_expansion}), we can
identify the macroscopic equations as follows:
\begin{eqnarray}
\label{macro_equations} \frac{d x(t)}{dt}&=& k_0 (x_0-x(t)-y(t)/K_v)
z_t -k_1 x(t) +k_2 y(t)/K_v -k_3 x(t) - k_4 c(t) x(t)
\\ \nonumber
\frac{d y(t)}{dt}&=& k_1 x(t) K_v -k_2 y(t)
\\ \nonumber
\frac{d c(t)}{dt}&=& k_8 y(t) -k_9 c(t) -k_4 c(t) x(t)
\\ \nonumber
\frac{d d(t)}{dt}&=& k_{10} y(t) -k_{11} d(t)
\\ \nonumber
\frac{d z(t)}{dt}&=& S \cdot k_6 -k_7 z(t) - k_5 d(t) z(t)
\end{eqnarray}
The above macroscopic equations are nothing but the deterministic
equations of the corresponding stochastic model in the limit of
infinitely large $\Omega$, i.e., in the limit of negligible
fluctuations.

The steady state solutions of the macroscopic equations can be obtained
by setting the left-handed sides of the Eq.~(\ref{macro_equations}):
$\bar{y}=k_1 K_v \bar{x}/k_2$, $\bar{c}=k_1 k_8 K_v \bar{x}
/(k_2 k_9)$, $\bar{z}=k_6 S/(k_7 + k_1 k_5 k_{10} K_v \bar{x}/(k_2
k_{11}))$. The steady state solutions of $\bar{x}$ are the roots of the
cubic equation, i.e., $d x_t /dt=0$ and only one of three roots are
positive real number and biologically feasible solution. We perform a
linear stability analysis by small perturbation of the dynamic system
from the feasible steady state solution, i.e., 
$\delta \vec{x}(t) \equiv \vec{x}(t)-\bar{\vec{x}}$. We rewrite the 
Eq.~\ref{macro_equations} in terms of this vector
$\delta \vec{x}(t)$: $d \delta \vec{x}(t) /dt = M \cdot \delta \vec{x} (t)$.
The stability of the steady state of the
Eq.~(\ref{macro_equations}) is solely determined by the Jacobian matrix $M$:
\\ $M=\left(
\begin{array}{ccccc}
M_{11} & M_{12} & M_{13} & 0      & M_{15}\\ M_{21} & M_{22} & 0
&0      & 0      \\ M_{31}      & M_{32} & M_{33} & 0      & 0      \\
0   & M_{42} & 0      & M_{44} & 0      \\ 0      & 0 & & M_{54} &
M_{55}
\end{array}
\right)$
\\ where $M_{11}=-k_0 \bar{z}-k_1-k_3-k_4 \bar{c}$,
$M_{12}=(k_2- k_0 \bar{z})/\sqrt{K_v}$, $M_{13}=-k_4 \bar{x}$,
$M_{15}=k_o (x_0 -\bar{x}-\bar{y}/K_v)$, $M_{21}=k_1 \sqrt{K_v}$,
$M_{22}=-k_2$, $M_{31}=-k_4 \bar{c}$, $M_{32}=k_8 \sqrt{K_v}$,
$M_{33}=-k_4 \bar{x} -k_9$, $M_{42}=k_{10} \sqrt{K_v}$,
$M_{44}=-k_{11}$, $M_{54}=-k_5 \bar{z}$, $M_{55}=-k_5 \bar{d} -k_7$.

\begin{figure}[htp]
\centering
\includegraphics[height=6cm,angle=0]{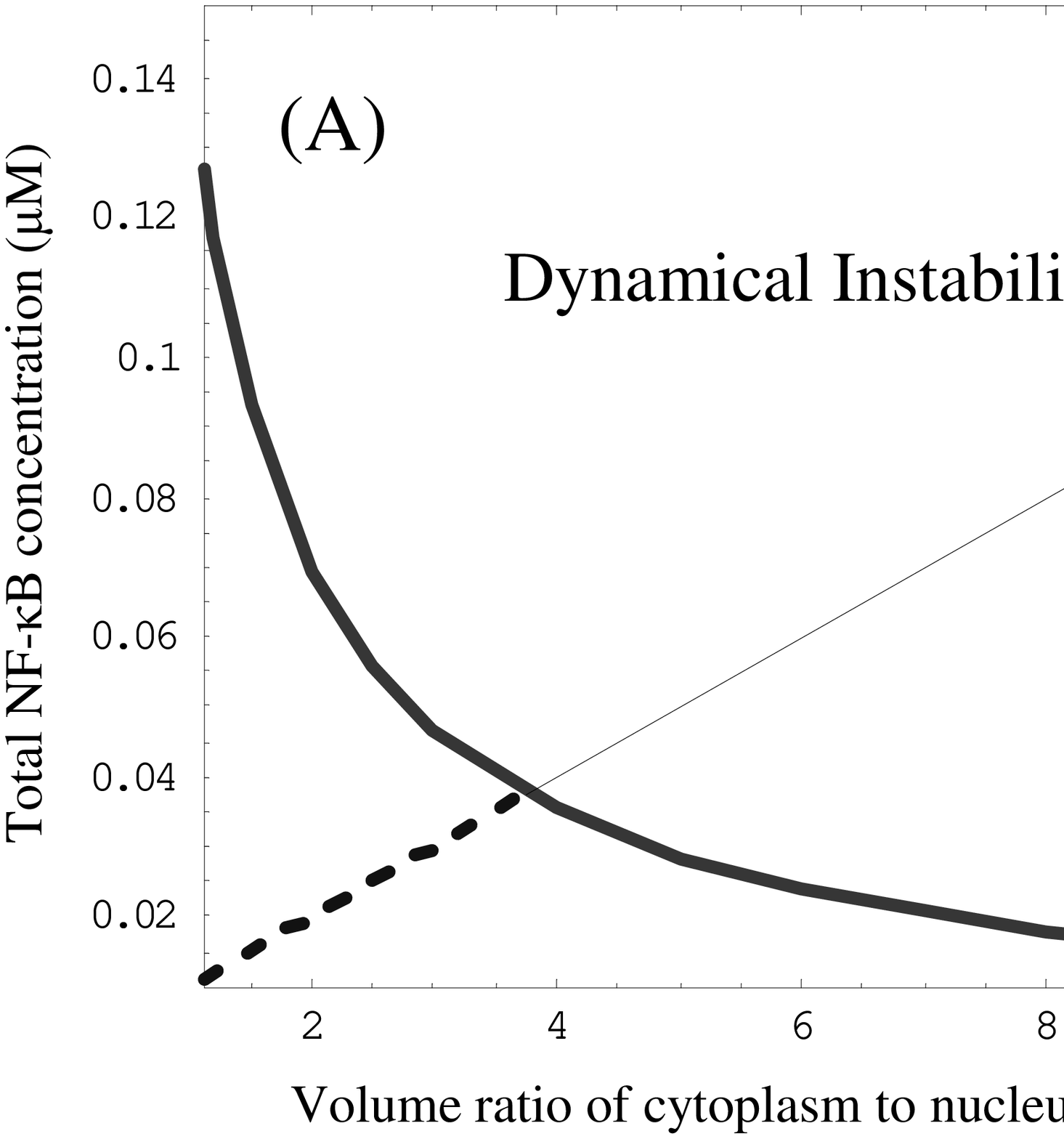}
\hfill
\includegraphics[height=6cm,angle=0]{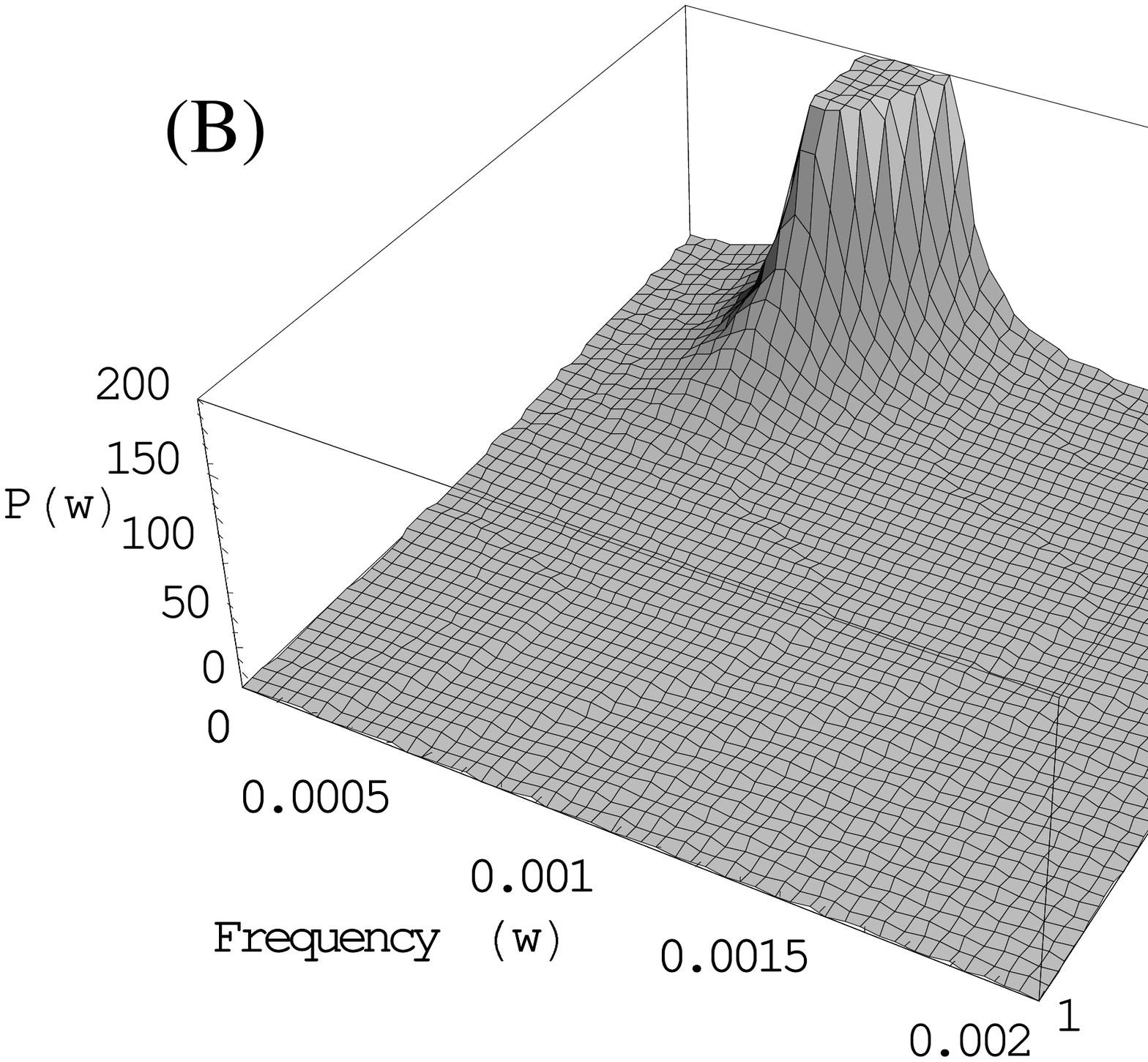}
\caption{Phase diagram of NF-$\kappa$ oscillation on two model parameter space, the volume ratio of cytoplasm to nucleus (Kv) and the total NFKB concetration, for the wild type case.
(A) shows the deterministic dynamical instability region where the fixed point of the
macroscopic equations is unstable due to the Hopf bifurcation. (B)
shows the power spectra in subcritical region along the diagonal
direction as indicated as the dashed line in (A). (B) shows the
power spectra at fixed value of Kv=2.5 in subcritical region. }
\label{fig2}
\end{figure}

The characteristic equation $| M - E \lambda |$ is the fifth order
polynomial equation and thus has five roots. The steady state of the
dynamical system is stable when the real parts of all the roots are
negative. Now for the macroscopic equations in
Eq. (\ref{macro_equations}), the three roots are always negative
real numbers within the biologically feasible domain of the kinetic
parameters under our consideration whereas the other two roots form
a pair of complex conjugate whose real part undergoes a sign change
as a function of the kinetic parameters. Thus, oscillation of this NF-$\kappa$B 
signaling network arise from Hopf bifurcation mechanism. 
A two-variable bifurcation diagram shown in
Fig.~\ref{fig2}(A) is constructed by numerically solving the
characteristic equation. When both total NF-$\kappa$B concentration and
volume ratio of cytoplasm to nucleoplasm increases along the
diagonal direction in Fig.~\ref{fig2}(A), the real parts of two
complex conjugate roots are changed from negative to positive values
and the onset of the Hopf bifurcation takes place at
$Kv^{Hopf}=3.754...$, giving rise to the emergence of a limit cycle.

\section{A linear Fokker-Planck Equation}
The next-to-leading order terms, $\Omega^{0}$, in the power
expansion of the master equation, give a partial differential
equation for the time evolution of the probability distribution
$\Pi_t (\underline{\xi})$, which has the form
\begin{equation}
\label{FP} \frac{\partial \Pi_t}{\partial t} =-\sum_{\alpha}
\frac{\partial A_{\alpha}(\underline{\xi}) \Pi_t}{\partial
\xi_{\alpha}} +\frac{1}{2} \sum_{\alpha \beta} B_{\alpha \beta}
\frac{\partial^2 \Pi_t}{\partial \xi_{\alpha} \partial \xi_{\beta}}
\end{equation}
where $A_{\alpha}(\underline{\xi})=\sum_{\beta} M_{\alpha \beta}
\xi_{\beta}$, $\alpha=X, Y, C, D, Z$, and $M$ is a Jacobian marix of
the Eq.~(\ref{macro_equations}) and $B$ are provided as below:
\\ $B=\left(
\begin{array}{ccccc}
B_{11} & B_{12} & B_{13} & 0      & 0      \\ B_{21} & B_{22} & 0
& 0      & 0      \\ B_{31}      & 0      & B_{33} & 0      & 0
\\ 0      & 0      & 0      & B_{44} & 0      \\ 0      & 0      &
& 0      & B_{55}
\end{array}
\right)$
\\  where the elements of both matrices $M$ and $B$ are
fully determined by the steady state solution of the macroscopic
equations, denoted by $\bar{\alpha}$: $B_{11}=k_o \bar{z}(
x_0-\bar{x}-\bar{y} /K_v )+k_1 \bar{x} +k_2 \bar{y}/K_v + k_3
\bar{x} + k_4 \bar{x} \bar{c}$, $B_{12}=B_{21}=-2 k_1 \sqrt{K_v}
\bar{x} - 2 k_2 \bar{y} / \sqrt{K_v}$, $B_{13}=B_{31}=2 k_4 \bar{x}
\bar{c}$, $B_{22}=k_2 \bar{y} + k_1 \bar{x} K_v$, $B_{33}=k_4
\bar{x} \bar{c}+k_8 \bar{y} + k_9 \bar{c}$, $B_{44}=k_{10} \bar{y} +
k_{11} \bar{d}$, $B_{55}=k_5 \bar{d} \bar{z} + k_6 S + k_7\bar{z}$.

The probability distribution at next-to-leading order is
completely determined by two $5 \times 5$ matrices. $A_{\alpha}
(\xi)$ remain linear functions of the $\xi_{\beta}$ and the
$B_{\alpha \beta}$ remain independent of them. Even though it is a
characteristic of a large $\Omega$ expansion for a single
compartmental model that $M$ is the Jacobian matrix which
determines the stability at a fixed point. The partial
differential equation is a linear Fokker-Planck equation, a
continuous version of the master equation and can be solved
exactly, given the initial condition that
$\Pi_{t=0}=\delta(\underline{\xi}-\underline{\xi_0})$. The
resulting solution of the probability distribution is a
multi-variate Gaussian (Van Kampen 2001).

The form of Eq.~(\ref{FP}) is not so useful for this purpose, but
fortunately there is a completely equivalent formulation of the
stochastic process which is ideally suitable for Fourier
transformations. The linear Fokker-Planck equation for the
probability distribution function is readily converted to a set of
stochastic differential equations of the Langevin type for the
actual stochastic variables $\xi$. The equivalent Langevin
equations are 
\begin{equation}
\label{Langevin}
\frac{d \vec{\xi}(t)}{dt}=M \cdot \vec{\xi}(t) +\vec{\eta}(t),
\end{equation}
where $\eta_i (t)$ is a Gaussian noise with zero mean and with a
correlation function given by
\begin{equation}
\label{noise-noise}
\langle \eta_{\alpha} (t) \eta_{\beta} (t') \rangle
= B_{\alpha \beta} \delta(t-t').
\end{equation}

To search for noise-induced oscillations, one of the most useful
diagnostic tools is the power spectrum $P(\omega)=\langle
|\tilde{\xi}(\omega)|^2 \rangle$, where $\tilde{\xi}(\omega)$ is the
Fourier transform of $\xi(t)$. Taking the Fourier transform of
Eq.~(\ref{Langevin}) gives
\begin{equation}
-i \omega \tilde{\xi}_i (\omega)= \sum_{j} M_{ij} \cdot \tilde{\xi}_j(\omega)
+ \tilde{\eta}_i (\omega).
\end{equation}
We obtain $\tilde{\xi}_i(\omega)=\sum_j \Phi_{ij}(i\omega)
\tilde{\eta}_j (\omega)$ where $\Phi_{ij}(i\omega)$ is the inverse
of $-i \omega \delta_{ij} -M_{ij}$. Averaging the squared modulus
of $\tilde{\xi}_i$ gives the power spectra
\begin{eqnarray}
P_i(\omega)=\langle |\tilde{\xi}_i (\omega)|^2 \rangle
&=&\sum_{jk} \Phi_{ij}(i \omega) B_{jk} \Phi^{\dagger}_{ki}(i \omega)
\end{eqnarray}
where we have used $\Phi^{\dagger}_{ij}(i \omega)=\Phi_{ji} (-i
\omega)$. Note that the power spectrum is completely determined by
two matrices $M$ and $B$ in Eq.~(\ref{Langevin}) and (\ref{noise-noise}).

\section{Power spectra and noise-induced oscillations}

In this section, we investigate the effect of noise on oscillatory pattern of nuclear NF-$\kappa$B component in wildt ype as well as I$\kappa$B knocked-out and A20 knocked-out mutants. All of our results are derived from the system size expansion and the small noise approximation and thus the presented results ae valid only when the system size is large and the fixed points of the macroscopic system are stable. 

In Fig. 2, we plot the power spectra as the volume ratio of cytoplasm to nucleus (Kv) changes from 1 to its bifurcation point, $Kv^{Hopf}=3.754...$ along the diagonal direction in two model parameter space. By this way, we satisfy the stability condition required for the validity of van kampen's method. 
Below the Hopf bifurcation point $Kv^{Hopf}$ where the deterministic dynamical system should be stable, the noise induces the oscillation of nuclear NF-$\kappa$B protein. The conditions for the exixtence of well-amplified NIO are (1) the peak of the power spectra is at non-zero resonant frequency and (2) the peak of the power spectra is much greater than its width which is approximated by the magnitude of the fluctuations of the protein copy number in the system. Clearly, these two conditions are satisfied in the power spectra in Fig. 2. In Fig. 2, non-zero resonant frequency where the power spectra has a peak is equivalent to the period of the stochastic time-series. When $Kv=Kv^{Hopf}$, the peak of the power spectra takes place at about $\omega=0.0005$ sec$^{-1}$ and thus the nuclear NF-$\kappa$B oscillates with about 3.5 hours of periodicity. As the volume ratio increases, the peak of the power spectra occurs at smaller frequency value and thus the period of the oscillations of NF-$\kappa$B protein concentration increases.

Now, we knock out one of two negative feedback loops, A20 or I$\kappa$B, and investigate if the oscillations of NF-$\kappa$B are robust in such a deletion. This deletion experiment provide us a clear idea about which negative feedback loop drive the NF-$\kappa$B oscillations. For this numerical experiments, all the original kintic rate constants in the NF-$\kappa$B system is kept unchanged while setting to zero the rate governing the mRNA I$\kappa$B synthesis for I$\kappa$B knocked-out mutant and the rate of the mRNA A20 synthesis for A20 knocked-out mutant.

In Fig. 3, even though I$\kappa$B negative feedback loop is absent, the overall qualitative dynamic features are invariant: the similar deterministic bifurcation curve as well as the similar power spectra. The only difference is that the deterministically dynamical instability domain is reduced slightly. But, the NIO of the NF-$\kappa$B is still observed just outside of the dynamical instability domain. But, in Fig. 4, when A20 negative feedack loop is knocked out, the entire phase plane is deterministically stable in the biologically feasible parameter domain. The inclusion of noise doesn't induce NIO of the NF-$\kappa$B.

\section{Conclusion}
We investigate the underlying mechanism of the biological oscillations of the NF-$\kappa$B signaling system by using a stochastic reduced order model and a small noise approximation. This analysis strongly hints that the noise-induced oscillation may be the source of the biological oscillations of NF-$\kappa$B in single cells. First, the oscillations of NF-$\kappa$B are driven by two negative feedback loops in deterministic dynamic system. Second, the stochastic fluctuations in the system induce the NF-$\kappa$B oscillations at the point where the deterministic system generates a stable dynamics because the noise can kick off the system away from its stable fixed point and the sytem takes a long excursive phase trajectory due to the closeby dynamic instability. The NIO of NF-$\kappa$B is robut against the deletion of I$\kappa$B negative feedback loop, but is not when A20 negative feedback loop is deleted. Thus, A20 may be essentially required for NF-$\kappa$B oscillations in single cells.

\begin{appendix}
\section{Appendix: Fast transition approximation between NF-$\kappa$B protein and DNA for transcription of I$\kappa$B and A20}
The five component model described in the transitions $T_8$ and $T_{10}$ in 
Eq.~(\ref{transition}) oversimplifies the very stochastic nature of 
transcription and translation, starting from $Y$ copy number of nuclear NF-$\kappa$B 
and ending with creation of a protein C or D, into a linear model. 
The mathematical description and its analysis of a DNA-protein interaction 
becomes far more complicated when the exclusiveness of the DNA operator 
site is taken into account, i.e, once a DNA operator site for a gene C 
is occupied, the remaining proteins can access no longer to the same 
operator site unless it is unoccupied. Here we attempt to incorporate 
the probabilitic operator transition into the model in Eq.~(\ref{transition}), 
by modifying especially two transitions, $T_8$ and $T_{10}$.

We briefly describe a probabilistic model of DNA-protein
interactions between nuclear NF-$\kappa$B and two I$\kappa$B and A20 promoters. A
Fig.~\ref{fig7} shows a transition between four different 
DNA operator states, i.e., four different combinations each of which 
consists of a state of I$\kappa$B promoter and one of A20 promoter. 
$P^{r,s}_t$ stands
for a joint probability of having a $r$ state of I$\kappa$B promoter and a
$s$ state of A20 promoter and $r$ and $s$ can take either 0
(unbounded) or 1 (NF-$\kappa$B-bound). We take a simplest assumption: (1) a
transition can change only the state of a single promoter. (2) The
change of the state of a promoter isn't influenced by and doesn't
affect the state of the other promoter. The transition from bounded
I$\kappa$B promoter state to I$\kappa$B promoter-free state is denoted by $q_0$
and transition from I$\kappa$B promoter-free to I$\kappa$B promoter-bound state
is by $q_1$ (likewise, $k_0$ and $k_1$ for A20). By stating that A20
mRNA synthesis rate is $\alpha_s$ and I$\kappa$B mRNA synthesis rate is
$\beta_r$ regardless of the A20 promoter state, we can write a time
evolution of a joint probability $P^{r,s}_t(C;D)$ of having the $C$
and $D$ copies of I$\kappa$B and A20 mRNA:
\begin{eqnarray}
\frac{d P^{r,s}_t (C,D)}{dt}&=&\alpha_s (P^{r,s}_t (C; D-1) -
P^{r,s}_t (C;D))+\beta_r (P^{r,s}_t (C-1;D)
\\ \nonumber
&-&P^{r,s}_t (C;D))+q_{r} P^{\hat{r},s}_t - q_{\hat{r}} P^{r,s}_t
+k_{s}P^{r,\hat{s}}_t - k_{\hat{s}} P^{r,s}_t
\end{eqnarray}
where $\hat{r}=(r+1)$ mod 2 and $\hat{s}=(s+1)$ mod 2. There are
four coupled equation like the above for the different combinations
of $(r,s)$, e.g., (0,0), (0,1), (1,0), (1,1).

We define four new variables (marginal probabilities) among which
only one variable will survive and the others will go extinct in the
limit of fast transition (which parameter?).
\begin{eqnarray}
W_t(C;D) &\equiv& \sum_{s,r} P^{r,s}(C;D) \\ \nonumber
X_t(C;D) &\equiv& k_1 (P^{0,0} +P^{1,0})-k_0 (P^{0,1}+P^{1,1}) \\ \nonumber
Y_t(C;D)  &\equiv& q_1 (P^{0,0} +P^{0,1})-q_0 (P^{1,0}+P^{1,1}) \\ \nonumber
Z_t(C;D) &\equiv& (k_1+q_1) P^{0,0} + (k_0 + q_0)P^{1,1}
-(k_o +q_1) P^{0,1} -(k_1+q_0) P^{1,0} \\ \nonumber
\end{eqnarray}
Here W, X, Y, and Z measures are four unique combinations which can
be made of two groups, each group being composed of a choice of two
original states. X, Y, and Z describes the inflow and the outflow
between those two groups denoted by the respective transition rates
and will be nullified In the long time limit because the steady
state flow between two chosen groups should be balanced. After the
substitution of $P^{r,s}$ with new variables $W, X, Y$, and Z, we
obtain time evolution equation of a new variable $W_t(C;D)$ as
follows:
\begin{eqnarray}
\label{master_equation_fast_operator}
\\ \nonumber
\frac{d W_t(C;D)}{dt}&=&\frac{q_0 \beta_0 + q_1 \beta_1 }{q_0 + q_1}
(W_t(C-1;D)-W_t(C;D)) + \frac{k_0 \alpha_0 + k_1 \alpha_1 }{k_0 +
k_1}  (W_t(C;D-1)-W_t(C;D))
\\ \nonumber
&+& \frac{\alpha_0 - \alpha_1 }{k_0 + k_1}  (X_t(C;D-1)-X_t(C;D))
+ \frac{\beta_0 - \beta_1 }{q_0 + q_1}  (Y_t(C-1;D)-Y_t(C;D))
\\ \nonumber
&=&\frac{q_0 \beta_0 + q_1 \beta_1 }{q_0 + q_1}  (W_t(C-1;D)-W_t(C;D))
+ \frac{k_0 \alpha_0 + k_1 \alpha_1 }{k_0 + k_1}  (W_t(C;D-1)-W_t(C;D))
\\ \nonumber
&+& \frac{k_0 k_1 (\alpha_0 - \alpha_1)^2 }{(k_0 + k_1)^3}
(W_t(C;D)-2W_t(C;D-1)+W_t(C;D-2))
\\ \nonumber
&+& \frac{q_0 q_1 (\beta_0 - \beta_1)^2}{(q_0 + q_1)^3}
(W_t(C;D)-2W_t(C-1;D)+W_t(C-2;D))
\end{eqnarray}
The time-evolution equations of the other three variables, X, Y, and
Z, rapidly approaches their respective quasi-steady states in the
limit of fast transition (we can show dX/dt =-$\gamma$X+... in the
limit of large $\gamma$ value.). The expressions in the last two
lines are obtained from setting the time evolution equations of X
and Y to zero and assuming the second derivatives of X, Y, and Z
with respect to C and D (See my notes for details).

Then, we will perform a small noise approximation of a master equation 
describing fast operator transition presented in 
Eq.~(\ref{master_equation_fast_operator}). We define three new transitions 
which decribe the association and the dissociation of the nuclear NF-$\kappa$B 
proteins with two distinct DNA promotors for gene I$\kappa$B and A20 and 
the DNA-state dependent protein synthesis. 
\begin{eqnarray}
\label{transition_fast_operator}
\delta_1 &=& \delta^{0}_1 \Omega_{N} (\frac{Y_t}{\Omega_N})[DNA]=\delta^{0}_1 Y_t
\\ \nonumber 
\delta_0 &=& \delta^{0}_0 \Omega_N [DNA]=\delta^{0}_0 \Omega_N
\\ \nonumber
\gamma_{1/0} &=& \gamma^{0}_{1/0} \Omega_C (\frac{RPII}{\Omega_N})[DNA]
=\gamma^{0}_{1/0} \Omega_C
\end{eqnarray}
where $\delta$ can be either $q$ for I$\kappa$B promotor or $k$ for A20 promotor 
and $\gamma$ can take either $\alpha$ for I$\kappa$B protein synthesis or 
$\beta$ for A20 protein synthesis whose transitions are depicted in Fig.~\ref{fig7}.
We separate the protein-DNA interactions from the actual transcription 
events which involve RNA polymerase II recruitement and transcriptional 
elongation, and mRNA synthesis, and protein production in Ribosome. 
Again, we lump two mRNA and protein syntheses into an effective event 
represented by $\gamma$ reaction for reduction of model dimension and 
simplicity of mathematical analysis. Here we assume a binary reaction 
between DNA promotor site and NF-$\kappa$B protein 
for $\delta_1$ and between RNA polymerase and DNA operator site for $\gamma$. 
Also, we assume a constant concentration of DNA promotor in nucleus and that 
of RNA polymerase in nucleus and absorb each into the corresponding reaction 
rate.

After substituting $Y_t$ with van Kampen's Ansatz and taking a continuum 
approximation of a step operator, we can rewrite 
the Eq.~(\ref{master_equation_fast_operator}) as follows:
\begin{eqnarray}
\label{master_equation_fast_operator_small_noise}
\frac{W_t(C;D)}{dt} &=&
\frac{\partial \omega_t(\xi_C;\xi_D)}{\partial t}
-\sum_{C,D}\Omega^{1/2}_C \frac{d \alpha}{dt} 
\frac{\partial \omega_t(\xi_C;\xi_D)}{\partial \xi_{\alpha}}
\\ \nonumber
&=&
-\Omega^{1/2}_C(A^D \frac{\partial}{\partial \xi_D}
+A^C \frac{\partial}{\partial \xi_C}) \omega_t(\xi_C;\xi_D)
\\ \nonumber
&+&\Omega^{0}_C 
(
\frac{\partial^2 B^D}{\partial \xi^2_D}
+\frac{\partial^2 B^C}{\partial \xi^2_C}
-\sqrt{K_v}\frac{\partial C^D}{\partial \xi_D}
-\sqrt{K_v}\frac{\partial C^C}{\partial \xi_C}
) \omega_t(\xi_C;\xi_D)
+O(\Omega^{-1/2})
\end{eqnarray}
where 
\begin{eqnarray}
A^D &=& \frac{k^0_0 \alpha^0_0+k^0_1 \alpha^0_1 y}{k^0_0 + k^0_1 y}
\\ \nonumber
B^D &=& A^D+\frac{k^0_0k^0_1 (\alpha^0_0 - \alpha^0_1)^2 y K_v}
{(k^0_0 + k^0_1 y)^3}
\\ \nonumber
C^D &=& \frac{k^0_1 \xi_Y}{k^0_0+k^0_1 y}(\alpha^0_1 
- \frac{k^0_0 \alpha^0_0+k^0_1 \alpha^0_1 y}{k^0_0 + k^0_1 y})
\\ \nonumber
A^C &=& \frac{q^0_0 \beta^0_0+q^0_1 \beta^0_1 y}{q^0_0 + q^0_1 y}
\\ \nonumber
B^C &=& A^C+\frac{q^0_0 q^0_1 (\beta^0_0 - \beta^0_1)^2 y K_v}
{(q^0_0 + q^0_1 y)^3}
\\ \nonumber
C^C &=& \frac{q^0_1 \xi_Y}{q^0_0+q^0_1 y}(\beta^0_1 
- \frac{q^0_0 \beta^0_0+q^0_1 \beta^0_1 y}{q^0_0 + q^0_1 y})
\end{eqnarray}
in the derivation, we use a taylor expansion of the transition rate in 
the limit of a large volume $\Omega_C$, e.g., 
$1/(1+\frac{k^0_1 \xi_Y \Omega^{-1/2}_C}{k^0_0 + k^0_1 y})^N
\sim 1-N \frac{k^0_1 \xi_Y \Omega^{-1/2}_C}{k^0_0 + k^0_1 y}$.

\end{appendix}

\newpage

\newpage

\newpage
%
%
%
%
%
\begin{figure}[htp]
\centering
\includegraphics[height=6cm,angle=0]{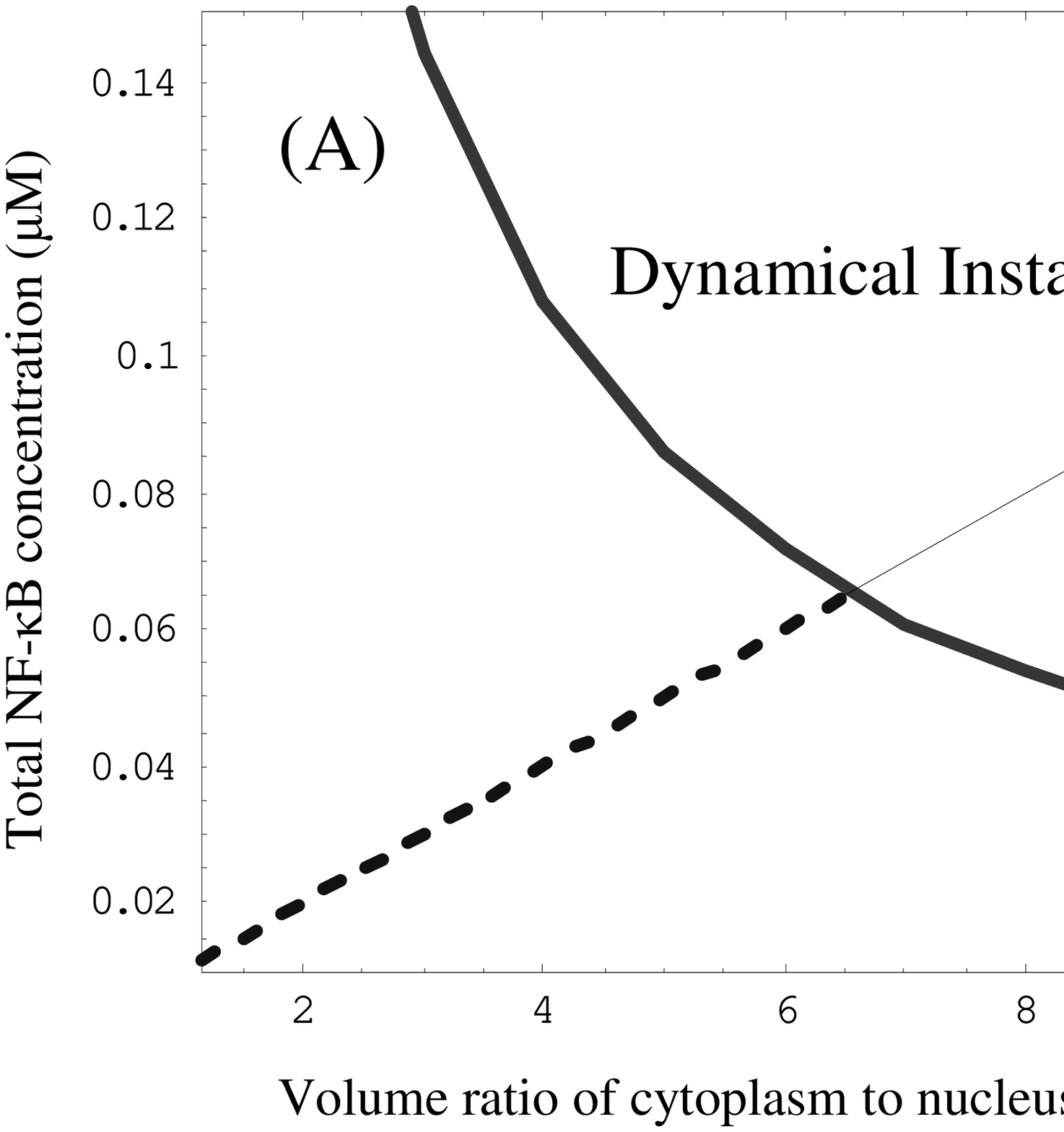}
\hfill
\includegraphics[height=6cm,angle=0]{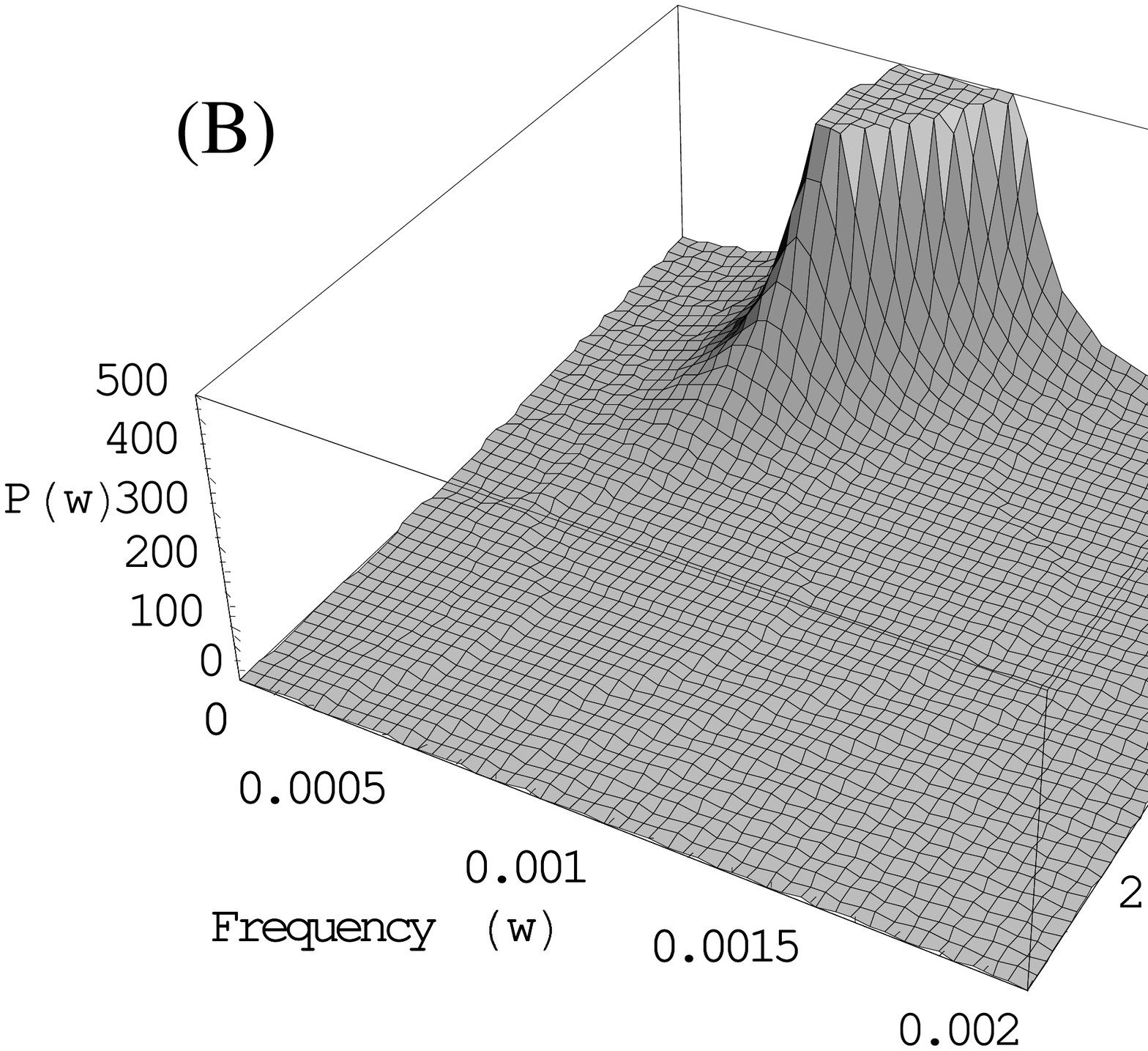}
\caption{Phase diagram and power spectra for the case of I$\kappa$B knocked-out mutant. (A) shows the phase diagram of NF-$\kappa$B oscillations.  (B) demonstrates the power spectra of nuclear NF-$\kappa$B along the diagonal direction in (A) (dashed line).} 
\label{fig5}
\end{figure}

\newpage

\begin{figure}[htp]
\centering
\includegraphics[height=6cm,angle=0]{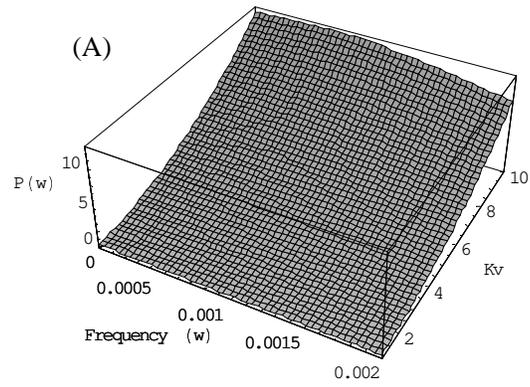}
\caption{Power spectra for the case of A20 knocked-out mutant. The phase diagram is not present because it contains a dynamically stable region. (A) shows the
power spectra along the diagonal direction from Kv=1 to Kv=10.}
 \label{fig6}
\end{figure}

\newpage

\begin{figure}[htp]
\centering
\includegraphics[height=10cm,angle=0]{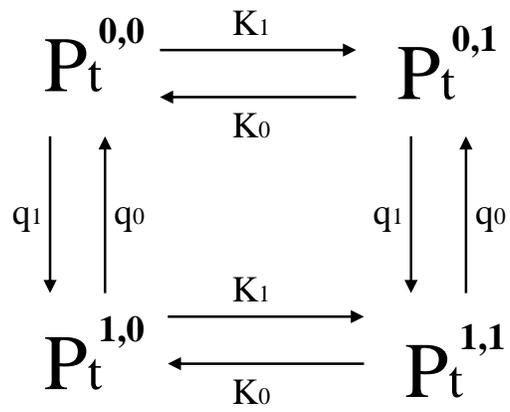}
\caption{Transition between four states consisting of I$\kappa$B and A20
promoters. $P^{r,s}_t$ stands for the probability having "r" state
of the I$\kappa$B promoter and "s" state of the A20 promoter. The
transition between two states is denoted by a transition rate,
either $q$ or $K$.} \label{fig7}
\end{figure}

\end{document}